\documentclass[10pt, journal]{IEEEtran}

\usepackage{pgfplots}
\usepackage{subfigure}
\usepackage{caption}
\usepackage{graphics}
\usepackage{xcolor}
\usepackage{balance}

\usepackage{tikz}
\usetikzlibrary{calc, math, arrows.meta}

\usepackage{textcomp}

\usepackage[noadjust]{cite}

\usepackage[T1]{fontenc}

\usepackage[cmintegrals]{newtxmath}

\usepackage{bm}
\usepackage{array}
\usepackage{url}

\newtheorem{lemma}{Lemma}

\begin{document}

\IEEEoverridecommandlockouts
\title{Spatio-Temporal Waveform Design in Active Sensing Systems with Multilayer Targets}
\author{
\IEEEauthorblockN{Ali Kariminezhad, \textit{Student Member, IEEE}, and Aydin Sezgin, \textit{Senior Member, IEEE}}\\
\thanks{
A. Kariminezhad and A. Sezgin are with the Institute of Digital Communication Systems, Ruhr-Universit\"at Bochum (RUB), Germany (emails: \{ali.kariminezhad, aydin.sezgin\}@rub.de).
}}
\maketitle
\pagestyle{empty}
\thispagestyle{empty}
\begin{abstract}
In this paper, we study the optimal spatio-temporal waveform design for active sensing applications. For this purpose a multi-antenna radar is exploited. The targets in the radar vision are naturally composed of multiple layers of different materials. Therefore, the interaction of these layers with the incident wave effects targets detection and classification. In order to enhance the quality of detection, we propose to exploit space-time waveforms which adapt with the targets multilayer response. We consider the backscattered signal power as the utility function to be maximized. The backscattered signal power maximization under transmit signal power constraint is formulated as a semidefinite program (SDP). First, we assume a single-target scenario, where the resulting SDP yields an analytical solution. Second, we study the optimal waveform which considers the angle uncertainties of a target in the presence of a clutter. Third, having multiple targets and multiple clutters, the weighted sum of the backscattered signals power from the targets is maximized to deliver the backscattered power region outermost boundary. We observe that, when the targets material is given, the backscattered signal power can be significantly increased by optimal spatio-temporal waveform design. Moreover, we observe that by utilizing multiple temporal dimensions in the waveform design process, the number of exploited antennas can be significantly decreased.
\end{abstract}

\maketitle

\section{Introduction}

Radio detection and ranging (radar) has been widely investigated for various applications including military, medical, driving assistant, traffic control systems and many others. However, recent applications of radar is not limited to only detection and ranging of the objects, but also involves material sensing and classification. The performance of radar is highly dependent on the transmit waveform design. This design can be performed in analog domain depending on the application, e.g., continuous-wave (CW) radar for ranging application. However, digital domain processing exploits the capability of radar for applications such as material response estimation. Mark R. Bell in his early work~\cite{Bell1993} has investigated the waveform design which maximizes the mutual information between received signal and the impulse response of a pointed target. Exploiting multi-antenna radar, the authors in~\cite{Leshem2007} studied the power allocation problem considering multiple extended targets. The authors in~\cite{Stoica2005,Ahmed2011,Lipor2014} study incident power maximization problem for multiple pointed targets given corresponding array response vectors. In those works, exploiting multiple antenna at the radar transmitter, the spatial domain covariance matrix is designed to enhance the incident signal power.

In this work, we put one step further and involve the interaction of multilayer material with the incident signal into account when designing the transmit waveform. The incident signal penetrates the multilayer material to a particular depth depending on the incident signal power and its frequency spectrum, however, the layers impedance attenuates the signal power. Moreover, a variant of the incident signal is reflected from the boundaries of the layers depending on the particular reflection coefficient of the layers. These phenomena are functions of the layers' magnetic permeability, dielectric permittivity and electrical conductivity. Hence, for the spatio-temporal waveform design, these parameters need to be known priori. The authors in~\cite{Duvillaret1996} propose a method to extract these parameters by terahertz time-domain spectroscopy. Hence, given these parameters, our goal is to maximize the backscattered signal power from the multilayer materials' surface for an improved parameter estimation, detection, classification, etc.. This type of waveform design can be in particular considered for medical imaging applications, where the materials of body organs, e.g, tissues, kidney, etc. are known priori.

First, we study the basic scenario with a single multilayer target. The backscattered signal power maximization problem under transmit power constraint is formulated as a semidefinite program. This problem is solved analytically. Second, we investigate a robust design problem which captures the targets position uncertainties. Third, we consider the scenario with multiple targets and multiple clutters. In order to determine the targets backscattered signal power outermost boundary, the weighted sum backscattered power maximization problem is formulated under clutters' backscattered power constraint and transmit power constraint. This problem is solved numerically by interior-point methods to deliver optimal spatio-temporal transmit covariance matrix~\cite{Boyd2004}. This covariance matrix is then utilized to obtain optimal beamforming and power allocation solutions jointly over space (antenna) and time. Exploiting this design for given targets material, we observe significant improvement in the backscattered signal power.

\section{System Model}
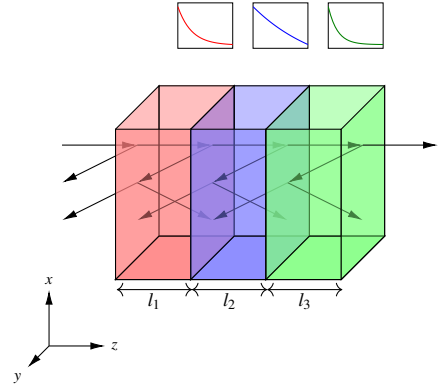
\begin{figure}
\centering
\tikzset{every picture/.style={scale=0.5}, every node/.style={scale=0.6}}
\begin{tikzpicture}[
			stylea/.style={fill = red!50, fill opacity = 0.5},
			styleb/.style={fill = blue!50, fill opacity = 0.5},
			stylec/.style={fill = green!50, fill opacity = 0.5},
			arrow/.style={-{Latex[length=2mm,width=1mm]}}
			]
			\def\h{4}
			\def\w{2}
			\def\d{3}
			
			\def\boxBack[#1]#2{
				\def\x{#2}
				\def\style{#1}
				
				\draw[\style] ({\x*\w}, {-\h/2}, {-\d/2}) -- +(0, \h) -- +(\w, \h) -- +(\w, 0) -- cycle;
				\draw[\style] ({\x*\w}, {-\h/2}, {\d/2}) -- +(0, 0, -\d) -- +(\w, 0, -\d) -- +(\w, 0) -- cycle;
				\draw[\style] ({\x*\w}, {-\h/2}, {\d/2}) -- +(0, \h) -- +(0, \h, -\d) -- +(0, 0, -\d) -- cycle; 
			}
		
			\def\boxFront[#1]#2{
				\def\x{#2}
				\def\style{#1}
				
				\draw[\style] ({\x*\w}, {-\h/2}, {\d/2}) -- +(0, \h) -- +(\w, \h) -- +(\w, 0) -- cycle;
				\draw[\style] ({\x*\w}, {-\h/2}, {\d/2}) -- +(0, 0, -\d) -- +(\w, 0, -\d) -- +(\w, 0) -- cycle;
				\draw[\style] ({(\x+1)*\w}, {-\h/2}, {\d/2}) -- +(0, \h) -- +(0, \h, -\d) -- +(0, 0, -\d) -- cycle; 
			}
			
			\coordinate (center) at (0, 1);
			
			\draw[arrow] (center) +(-2, 0) -- +(0, 0); 
			\draw[arrow] (center) +(0, 0) -- +(-2, -1); 
			\draw[arrow] (center) ++(0, -1) -- +(-2, -1); 
			\boxBack[stylea]{0}
			\draw[arrow] (center) +(0, 0) -- +(\w, 0); 
			\draw[arrow] (center) +(\w, 0) -- +(0, -1); 
			\draw[arrow] (center) +(0, -1) -- +(\w, -2); 
			\draw[arrow] (center) ++({\w*1}, -1) -- +(-2, -1); 
			\boxFront[stylea]{0}

			\boxBack[styleb]{1}
			\draw[arrow] (center) +(\w, 0) -- +({\w*2}, 0); 
			\draw[arrow] (center) ++({\w*2}, 0) -- +(-\w, -1); 
			\draw[arrow] (center) ++(\w, -1) -- +(\w, -1); 
			\draw[arrow] (center) ++({\w*2}, -1) -- +(-2, -1); 
			\boxFront[styleb]{1}

			\boxBack[stylec]{2}
			\draw[arrow] (center) +({\w*2}, 0) -- +({\w*3}, 0);
			\draw[arrow] (center) ++({\w*3}, 0) -- +(-\w, -1); 
			\draw[arrow] (center) ++({\w*2}, -1) -- +(\w, -1); 
			\boxFront[stylec]{2}
			
			\draw [arrow](center) ++({\w*3}, 0) -- +(2, 0);
			
			
			\foreach \plotIndex in{0}{
							\coordinate (plot\plotIndex) at ({\w*\plotIndex + 0.5}, 3, {-\d/2});
							\begin{axis}[
									ticks=none,
									compat = 1.11,
									width = 4.5cm,
									height = 4cm,
									at = {(plot\plotIndex)},
									xmin = 0,
									xmax = 4,
									domain = 0:4,
								]
									\addplot+[no marks, red, thick]{exp(-x)};
							\end{axis}
						}
          \foreach \plotIndex in{1}{
							\coordinate (plot\plotIndex) at ({\w*\plotIndex + 0.5}, 3, {-\d/2});
							\begin{axis}[
									ticks=none,
									compat = 1.11,
									width = 4.5cm,
									height = 4cm,
									at = {(plot\plotIndex)},
									xmin = 0,
									xmax = 4,
									domain = 0:4,
								]
									\addplot+[no marks, thick]{exp(-0.2*x)};
							\end{axis}
						}						
          \foreach \plotIndex in{2}{
							\coordinate (plot\plotIndex) at ({\w*\plotIndex + 0.5}, 3, {-\d/2});
							\begin{axis}[
									ticks=none,
									width = 4.5cm,
									height = 4cm,
									at = {(plot\plotIndex)},
									xmin = 0,
									xmax = 4,
									domain = 0:4,
								]
									\addplot+[no marks,green!50!black,thick]{exp(-1.5*x)};
							\end{axis}
						}		
						
			\coordinate (cord) at (-1.7, -4, -3);	
			\draw [<->] (cord) -- +(2, 0, 0)
			node at (-0.7,-4.3,-3) {\large$l_1$};
			\coordinate (cord) at (0.3, -4, -3);
			\draw [<->] (cord) -- +(2, 0, 0)
			node at (1.3,-4.3,-3) {\large$l_2$};
			\coordinate (cord) at (2.3, -4, -3);
			\draw [<->] (cord) -- +(2, 0, 0)
			node at (3.3,-4.3,-3) {\large$l_3$};
			
			\coordinate (cord) at (-3.5, -5.5, -3);
			\draw [arrow] (cord) -- +(1.5, 0, 0) node[right]{$z$};
			\draw [arrow] (cord) -- +(0, 1.5, 0) node[above]{$x$};
			\draw [arrow] (cord) -- +(0, 0, 1.5) node[below left]{$y$};
		\end{tikzpicture}
\caption{Three layer structure. The incident signal passes through the multilayer material, attenuates at each layer exponentially and is reflected from the boundaries between layers.}
\label{fig:SystemModel}
\end{figure}

We consider an active sensing system, where a multi-antenna radar is deployed to detect or identify unknown parameters of targets with multiple layers~Fig~\ref{fig:SystemModel}. Initially, we consider a scenario with a single multilayer target in this section.
Let the incident signal at the target surface in time instant $t$ be
\begin{align}
r_{\text{inc}}(t)=\mathbf{a}^H_{\theta}\mathbf{s}(t),\label{eq:SysModel}
\end{align}
where the transmit signal from a transmitter equipped with $M$ antennas at time instant $t$ is represented by $\mathbf{s}(t)\in\mathbb{C}^M$. Moreover, the array response vector corresponding to the static target is given by $\mathbf{a}_{\theta}\in\mathbb{C}^M$. Suppose that the target is located at the observation angle $\theta$, then the transmit array response vector is given by
\begin{align}
\mathbf{a}_{\theta}=\begin{bmatrix}
1 & e^{-j\frac{2\pi D}{\lambda}\sin\theta} & \cdots & e^{-j\frac{2(M-1)\pi D}{\lambda}\sin\theta}
\end{bmatrix}^{T},
\end{align}
where the first antenna (nearest antenna to the target) is considered as the reference antenna.

Now, suppose that the multilayer target has $L$ layers. As the incident signal penetrates the multilayer material in z-direction,~Fig.~\ref{fig:SystemModel}, a portion of the incident signal power dissipates as
\begin{align}
P_\text{d}(z)=P_{r_\text{inc}}e^{-\beta_i z},
\end{align}
where $P_{r_\text{inc}}$ is the incident signal power. Notice that $\beta_i$ denotes the layer $i$th signal attenuation factor. Let $l_i$ be the physical depth of $i$th layer. Then, the signal reached at the boundary of $i-1$th and $i$th layer has the power
\begin{align}
P^{(i)}_{r_\text{inc}}=&P^{(i-1)}_{r_\text{inc}}\int_{0}^{l_i}e^{-\beta_i z}dz\nonumber\\
=&\frac{1}{\beta_i}P^{(i-1)}_r\left(1-e^{-\beta_i l_i}\right),\quad \forall i\in\mathcal{L}=\{1,...,L\}.
\end{align}
Here, we define the power attenuation at layer $i$ by $\alpha_i=\frac{1}{\beta_i}\left(1-e^{-\beta_i l_i}\right)$. Hence, the received signal at layer $i$ can be modeled as
\begin{align}
r_{\text{inc}}^{(i)}(t)=\prod_{q=1}^{i-1}
\sqrt{\alpha_q}r_{\text{inc}}(t),\quad \forall i\in\mathcal{L}.
\end{align}
This model captures the attenuation of the signal through the target till layer $i$. However, another portion of the signal is reflected back and forth at the boundaries between the layers. The reflection coefficient at the boundary of $i-1$th and $i$th layers is given by
\begin{align}
\rho_i=\frac{\eta_{i}-\eta_{i-1}}{\eta_{i}+\eta_{i-1}},\quad \forall i\in\mathcal{L},
\end{align}
where $\eta_i$ is the complex-valued impedance of $i$th layer. This impedance is given by
\begin{align}
\eta_i=\sqrt{\frac{\mu_i}{\epsilon_i}}\left(1-j\frac{\sigma_i w}{\epsilon_i}\right)^{-0.5},\forall i\in\mathcal{L},
\end{align}
where $\mu_i$, $\epsilon_i$ and $\sigma_i$ represent the magnetic permeability, dielectric permittivity and electrical conductivity of $i$th layer, respectively. Moreover, the angular frequency of the incident signal is denoted by $w$. Notice that, $\eta_i=\sqrt{\frac{\mu_i}{\epsilon_i}}$ assuming an ideal dielectric layer. 

Now, the signal reflected from the surface of the target is denoted by $\rho_1r_{\text{inc}}(t)$. Due to the attenuation of the signal through the first layer i.e., $\alpha_1$, the reflected signal from the boundary of the first and second layer is $\sqrt{\alpha_1}\rho_2r_{\text{inc}}(t)$. Ignoring multiple reflections between the boundaries for simplicity, the received signal from the boundary of the first and second layer at the surface of the target becomes $\alpha_1\rho_2r_{\text{inc}}(t)$. Thereby, the received signal from the boundary of $i-1$th layer and $i$th layer at the target surface is $\prod_{q=1}^{i-1}\alpha_q\rho_ir_{\text{inc}}(t)$.

Here, we assume that the reflected signal from each layer appears at the target surface at successive time instants with period $T$. For instant, the reflected signal at the first time instant only includes the reflection from the target surface, however the signal at the second time instant involves both the reflected signal from the surface and the first boundary. This can be validated by assuming equal travel-time multilayer objects~\cite{Orfanidis1988}. By this assumption, we model this multilayer target as a time-varying transfer function as
\begin{align}
g(t)= \sum_{i=1}^{L}\zeta_i \delta\Big(t-(i-1)T\Big),
\end{align}
where $\zeta_i=\prod_{q=1}^{i-1}\alpha_q\rho_i$ and $\delta(t)$ is the Dirac delta function. The reflected signal from each layer appears at the surface of the material to form the backscattered signal from the target. This signal is given by
\begin{align}
r_{\text{bsc}}(t)=&r_{\text{inc}}(t)\circledast g(t)=\\
=&\sum_{i=1}^{L}\zeta_i\int_{\tau=-\infty}^{\infty}r_{\text{inc}}(\tau)
\delta\Big(t-\tau-(i-1)T\Big)d\tau\nonumber\\
=&\sum_{i=1}^{L}\zeta_ir_{\text{inc}}\Big(t-(i-1)T\Big),
\label{eq:Conv1}
\end{align}
where $\circledast$ denotes the convolution operator. Sampling the backscattered signal at the sampling frequency $f_s=\frac{1}{T}$ gives
\begin{align}
r_{\text{bsc}}(n)=&\int_{\tau=-\infty}^{\infty}r_{\text{bsc}}(\tau)\delta(\tau-nT)d\tau\\
=&\sum_{i=1}^{L}\zeta_ir_{\text{inc}}\Big((n-i+1)T\Big).
\end{align}
Now by stacking $N$ time samples of $r_{\text{bsc}}(n)$ in a vector, we obtain
\begin{align}
\mathbf{r}_{\text{bsc}}=\mathbf{Z}\mathbf{A}_{\theta}\tilde{\mathbf{s}},
\end{align}
where
\begin{align}
\mathbf{Z}=&
  \begin{bmatrix}
  \zeta_1 & 0& 0&\cdots&0\\
  \zeta_2&\zeta_1& 0& \cdots&0\\
   \vdots \\
  \zeta_L & \zeta_{L-1} & \zeta_{L-2} & \cdots & 0\\
  0 & \zeta_L & \zeta_{L-1} & \cdots & 0\\
  \vdots \\
   \mathbf{0}^{T}_{N-L} & \zeta_L  & \zeta_{L-1}& \cdots & \zeta_1 \\
  \end{bmatrix}_{N\times N},\\ \nonumber \\
  \mathbf{A}_{\theta}=&[\mathbf{I}_N\otimes\mathbf{a}^{H}_{\theta}]_{N\times MN},
\end{align}
where $\otimes$ denotes the Kronecker product and $\mathbf{I}_N$ is the $N$-dimensional identity matrix and $\mathbf{0}_{N-L}$ is a vector of $N-L$ zeros. Moreover, the space-time transmit signal vector $\tilde{\mathbf{s}}\in\mathbb{C}^{MN}$ is given as
\begin{align}
\tilde{\mathbf{s}}=&\begin{bmatrix}
  \mathbf{s}(0) & ... &\mathbf{s}(N-1) 
  \end{bmatrix}^H,
\end{align}
where $\mathbf{s}(n)= \int_{\tau=-\infty}^{\infty}\mathbf{s}(\tau)\delta(nT-\tau)d\tau$. The space-time transmit signal is constructed as
\begin{align}
\tilde{\mathbf{s}}=\mathbf{U}\mathbf{w},
\end{align}
where $\mathbf{U}\in\mathbb{C}^{MN\times MN}$ and $\mathbf{w}\in\mathbb{C}^{MN}$ are the transmit space-time precoding matrix and information symbol vector, respectively. Here, we allow $\mathbb{E}\{\mathbf{w}\mathbf{w}^H\}=\mathbf{I}_{MN}$. Thus, the transmit signal power is embedded in the space-time transmit precoding matrix. This precoding matrix is intended to be designed optimally for backscattered signal power maximization purpose.

In the next section, we consider a scenario with single multilayer target. 

\section{Single Target}
Now the transmit signal precoding matrix needs to be optimized in order to maximize the backscattered signal power. The backscattered signal power is given by

\begin{align}
P_{\mathbf{r}_{\text{bsc}}}=\mathbb{E}\{\text{Tr}\left(\mathbf{r}_{\text{bsc}}\mathbf{r}_{\text{bsc}}^H\right)\}
=&\mathbb{E}\{\text{Tr}\left(\mathbf{Z}\mathbf{A}_{\theta}\tilde{\mathbf{s}}\tilde{\mathbf{s}}^H\mathbf{A}^H_{\theta}\mathbf{Z}^H\right)\}\nonumber\\
=&\text{Tr}\left(\mathbf{A}^H_{\theta}\mathbf{Z}^H\mathbf{Z}\mathbf{A}_{\theta}\mathbf{U}\mathbf{U}^H\right),\nonumber
\end{align}
where $\mathbb{E}\{.\}$ and $\text{Tr}(.)$ denote the expectation and trace operators, respectively. Defining   $\tilde{\mathbf{Z}}_{\theta}=\mathbf{A}^H_{\theta}\mathbf{Z}^H\mathbf{Z}\mathbf{A}_{\theta}$, the backscattered signal power is rewritten as
$
P_{\mathbf{r}_{\text{bsc}}}=\text{Tr}\left(\tilde{\mathbf{Z}}_{\theta}\mathbf{C}\right),
$
where $\mathbf{C}=\mathbf{U}\mathbf{U}^H$ is the transmit signal covariance matrix. The backscattered signal power maximization is then formulated as
\begin{subequations}\label{OP:A1}
\begin{align}
\max_{\mathbf{C}}\quad &\text{Tr}\left(\tilde{\mathbf{Z}}_{\theta}\mathbf{C}\right)
\tag{\ref{OP:A1}}\\
\text{s.t.}\quad & \text{Tr}\left( \mathbf{C}\right)\leq P_{\text{max}},\label{eq:PowerCons1}\\
&\mathbf{C}\succeq 0,\label{eq:PsdCons1}\\
&\mathbf{C}\in\mathcal{H}^{MN\times MN},\label{eq:HermitCons1}
\end{align}
\end{subequations}
where the transmit power constraint is denoted by~\eqref{eq:PowerCons1} with $P_{\text{max}}$ as the transmit power budget. The constraints~\eqref{eq:PsdCons1} and~\eqref{eq:HermitCons1} represent the properties of a covariance matrix to be positive semidefinite and Hermitian, respectively. Notice that $\mathcal{H}^{MN\times MN}$ denotes the cone of Hermitian matrices of dimension $MN\times MN$. Problem~\eqref{OP:A1} is a semidefinite program (SDP).
\begin{lemma}
The optimization problem~\eqref{OP:A1} has a single-rank solution, which is
\begin{align}
\mathbf{C}^{\star}=P_\text{max}\mathbf{v}_{<\text{max}>}\mathbf{v}^{H}_{<\text{max}>},
\end{align}
where $\mathbf{v}_{<\text{max}>}$ is the eigen-vector corresponding to the maximum eigen-value of $\tilde{\mathbf{Z}}_{\theta}$.

\textit{Proof}: Let the eigen-value decomposition of $\tilde{\mathbf{Z}}_{\theta}$ be $\tilde{\mathbf{Z}}_{\theta}=\mathbf{V}\boldsymbol\Lambda\mathbf{V}^{H}$, where $\mathbf{V}=[\mathbf{v}_1,\cdots,\mathbf{v}_N]$ is the eigen-vector matrix with the eigen-vector $\mathbf{v}_i$ in the $i$th column. Moreover, $\boldsymbol{\Lambda}$ is the diagonal matrix of eigen-values in increasing order, i.e., $\boldsymbol{\Lambda}=\text{diag}\big([\lambda_1,\cdots,\lambda_N]\big)$. Then optimization problem~\eqref{OP:A1} is reformulated as
\begin{subequations}\label{OP:A2}
\begin{align}
\max_{\mathbf{C}}\quad \sum_{i=1}^{N}\lambda_i\text{Tr}\big(\mathbf{v}_i\mathbf{v}^{H}_i\mathbf{C}\big)
\tag{\ref{OP:A2}}
\quad\quad \text{s.t.}\ \ \eqref{eq:PowerCons1}-\eqref{eq:HermitCons1}.
\end{align}
\end{subequations}
Notice that the eigen vectors of the optimal solution, i.e., $\mathbf{C}$, span the eigen directions of $\tilde{\mathbf{Z}}_{\theta}$. Hence,
$\mathbf{C}(p_1,\cdots,p_N)=\sum_{i=1}^{N}p_i\mathbf{v}_i\mathbf{v}^{H}_i$, where $p_i$ is the power dedicated for $i$th eigen direction. Now, problem~\eqref{OP:A2} can be reformulated as
\begin{subequations}\label{OP:A21}
\begin{align}
\max_{p_1,...,p_N}\quad \sum_{i=1}^{N}\lambda_i p_i
\tag{\ref{OP:A21}}
\quad\quad \text{s.t.}\ \  \sum_{i=1}^{N}p_i\leq P_{\text{max}},\label{eq:PowerCons21}
\end{align}
\end{subequations}
which is a linear program (LP) and can be solved by the simplex method~\cite{Antoniou2007}. In this program, the optimal solution lies on the simplex vertices with the value $\lambda_i$ at $i$th vertex. Hence, the optimal solution is at the vertex with the maximum value, i.e., $\lambda_N$. Moreover, since the affine objective function is maximized, the inequality constraint~\eqref{eq:PowerCons21} holds with equality at the optimal solution. This way the optimal power allocation solution is $[p^{\star}_1,\cdots,p^{\star}_N]=[0,\cdots,P_\text{max}]$ and $\mathbf{C}^{\star}=P_\text{max}\mathbf{v}_N\mathbf{v}^{H}_N$. 

In what follows, we study the robust waveform design when the target position is known partially.
\end{lemma} 
	\begin{figure*}
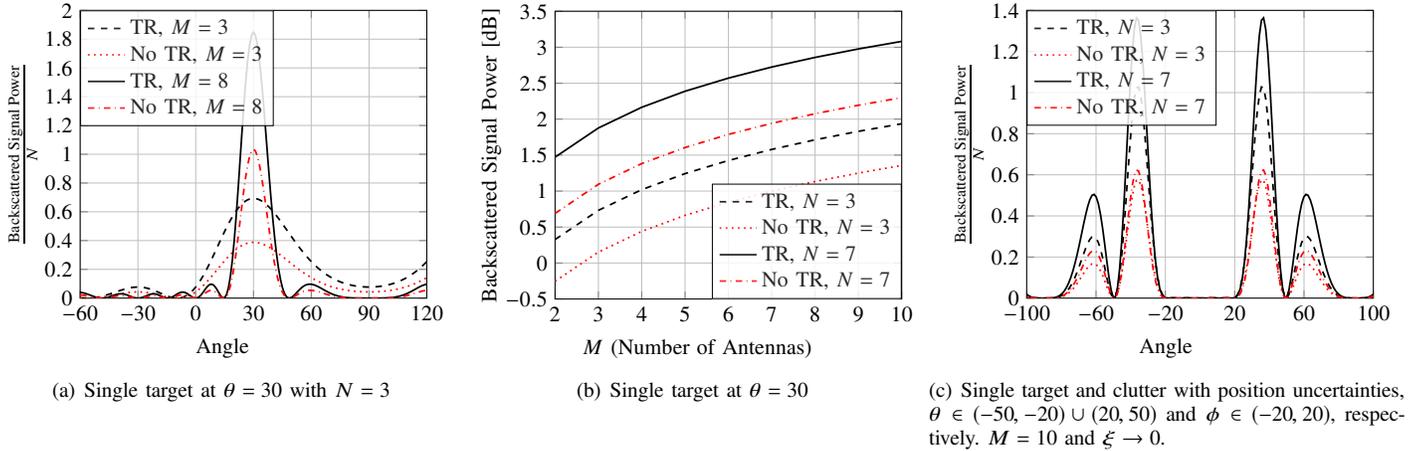

	\hspace*{-2cm}
			\centering
			\begin{minipage}[b]{0.25\textwidth}
			\subfigure[\footnotesize Single target at $\theta=30$ with $N=3$]{
				\tikzset{every picture/.style={scale=0.82}, every node/.style={scale=1}}
				\input{SingleTarget}
				\label{fig:SingleTarget}
	}
			\end{minipage}\quad\quad\quad\quad\quad
			\begin{minipage}[b]{0.25\textwidth}
			\subfigure[\footnotesize Single target at $\theta=30$]{
				\tikzset{every picture/.style={scale=0.82}, every node/.style={scale=1}}
				\begin{tikzpicture}

\begin{axis}[%
xmin=2,
xmax=10,
xlabel={$M$ (Number of Antennas)},
xmajorgrids,
xtick={2,3,4,5,6,7,8,9,10},
ymin=-0.5,
ymax=3.5,
ylabel={ Backscattered Signal Power [dB]},
ylabel near ticks,
ymajorgrids,
ytick={-0.5,0,0.5,1,1.5,2,2.5,3,3.5},
legend style={at={(axis cs:10,-0.5)},anchor= south east, draw=black,fill=white, fill opacity=0.85,legend cell align=left}
]
\addplot [color=black,dashed, thick]
  table[row sep=crcr]{2	0.326540212428036\\
  3	0.732005326297153\\
  4	1.01968740170313\\
  5	1.2428309543772\\
  6	1.42515250467433\\
  7	1.57930317410967\\
  8	1.71283458568477\\
  9	1.83061761862421\\
  10 1.93597812116706\\
  };
\addlegendentry{TR, $N=3$};

\addplot [color=red,dotted, thick]
  table[row sep=crcr]{2	-0.253602761427201\\
  3	0.151862345064373\\
  4	0.439544417795578\\
  5	0.662687969323319\\
  6	0.845009523613832\\
  7	0.999160209095015\\
  8	1.1326915994607\\
  9	1.25047462987975\\
  10 1.35583515310075\\
  };
\addlegendentry{No TR, $N=3$};
  
  \addplot [color=black,solid, thick]
    table[row sep=crcr]{2	1.47162235259914\\
    3	1.877087460056\\
    4	2.16476952403067\\
    5	2.38791307970043\\
    6	2.57023464085473\\
    7	2.72438530456746\\
    8	2.85791669609671\\
    9	2.97569973888962\\
    10	3.08106025780643\\
    };
  \addlegendentry{TR, $N=7$};

  \addplot [color=red,dashdotted, thick]
    table[row sep=crcr]{2	0.689139154683742\\
    3	1.0946042607627\\
    4	1.38228631884352\\
    5	1.60542988658371\\
    6	1.78775144721797\\
    7	1.94190212542463\\
    8	2.07543351800538\\
    9	2.19321655258895\\
    10	2.29857706724007\\
    };
  \addlegendentry{No TR, $N=7$};
  
      

\end{axis}
\end{tikzpicture}%
				\label{fig:SpatialTemporalTradeoff}
	}
			\end{minipage}\quad\quad\quad\quad\quad
            \begin{minipage}[b]{0.25\textwidth}
				\subfigure[\footnotesize Single target and clutter with position uncertainties, $\theta\in (-50,-20)\cup (20,50)$ and $\phi\in (-20,20)$, respectively. $M=10$ and $\xi\rightarrow 0$.]{
					\tikzset{every picture/.style={scale=0.82}, every node/.style={scale=1}}
					\input{SingleTargetCoverRegionWithClutterCoverRegion}
					\label{fig:SingleTargetCoverRegionWithClutterCoverRegion}
				
   }
				\end{minipage}		
			\caption{Backscattered signal power as a function of number of antennas and number of temporal dimensions utilized in the waveform design phase.}
		\end{figure*}

\subsection{Target Position Uncertainties}
Due to the fact that, a small error in the target positioning process yields tremendous reduction in the target backscattered signal power, we study the scenario with single multilayer target in an uncertainty angle $\theta\in (\theta_{\text{low}},\theta_{\text{high}})$ and single clutter in an uncertainty angle $\phi\in (\phi_{\text{low}},\phi_{\text{high}})$. For such a scenario, we investigate a waveform which guarantees an arbitrarily small backscattered signal power from the clutter uncertainty region, meanwhile, maximizing the target backscattered signal power. This design is conducted in a max-min optimization framework. Here, we define the sets $\mathcal{U}_\text{T}=\{u_{\text{T}_1},\cdots,u_{\text{T}_R}\}$ and $\mathcal{U}_\text{C}=\{u_{\text{C}_1},\cdots,u_{\text{C}_{R^{'}}}\}$. Moreover, we define $u_{\text{T}_1}=\theta_{\text{low}}$, $u_{\text{T}_R}=\theta_{\text{high}}$, $u_{\text{T}_j}-u_{\text{T}_{j-1}}=\frac{\theta_{\text{high}}-\theta_{\text{low}}}{R-1},\ \forall j\in\{2,\cdots,R\}$, and $u_{\text{C}_1}=\phi_{\text{low}}$, $u_{\text{C}_{R^{'}}}=\phi_{\text{high}}$, $u_{\text{C}_i}-u_{\text{C}_{i-1}}=\frac{\phi_{\text{high}}-\phi_{\text{low}}}{R^{'}-1},\ \forall i\in\{2,\cdots,R^{'}\}$. Now, we formulate the backscattered power maximization problem under clutters backscattered power constraint in an uncertainty region as
\begin{subequations}\label{OP:A3}
\begin{align}
\max_{\mathbf{C}}\quad &\min_{u_{\text{T}_i}\in\mathcal{U}_\text{T}}\ \text{Tr}\left(\tilde{\mathbf{Z}}_{u_{\text{T}_i}}\mathbf{C}\right)
\tag{\ref{OP:A3}}\\ 
\text{s.t.}\quad & \text{Tr}\left(\mathbf{A}^{H}_{u_{\text{C}_i}}\mathbf{A}_{u_{\text{C}_i}}\mathbf{C}\right)\leq \xi,\quad \forall i\in\{1,\cdots,R^{'}\}\\
& \eqref{eq:PowerCons1}-\eqref{eq:HermitCons1},
\end{align}
\end{subequations}
where $\tilde{\mathbf{Z}}_{u_{\text{T}_i}}=\mathbf{A}_{u_{\text{T}_i}}^H\mathbf{Z}^H\mathbf{Z}\mathbf{A}_{u_{\text{T}_i}}$. Notice that $\mathbf{A}_{u_{\text{T}_i}}=\mathbf{I}_N\otimes\mathbf{a}^{H}_{u_{\text{T}_i}}$. Moreover $\xi$ denotes the backscattered signal power constraint from the clutter which is located in the angle interval $(\phi_{\text{low}},\phi_{\text{high}})$.\\
By defining the auxiliary variable $\Gamma=\min_{u_{\text{T}_i}\in\mathcal{U}_\text{T}}\ \text{Tr}\left(\tilde{\mathbf{Z}}_{u_{\text{T}_i}}\mathbf{C}\right)$, problem~\eqref{OP:A3} is reformulated as

\begin{subequations}\label{OP:A4}
\begin{align}
\max_{\Gamma,\mathbf{C}}\quad & \Gamma
\tag{\ref{OP:A4}}\\
\text{s.t.}\quad & \Gamma\leq \text{Tr}\left(\tilde{\mathbf{Z}}_{u_{\text{T}_i}}\mathbf{C}\right),\quad \forall j\\
&\text{Tr}\left(\mathbf{A}^{H}_{u_{\text{C}_i}}\mathbf{A}_{u_{\text{C}_i}}\mathbf{C}\right)\leq \xi,\quad \forall i\\
&\eqref{eq:PowerCons1}-\eqref{eq:HermitCons1},
\end{align}
\end{subequations}
which is a SDP and can be solved efficiently by interior-point methods~\cite{Boyd2004}. So far, we have investigated single target scenario. 

In the next section we consider multiple targets and multiple clutters in the radar vision.

\section{Multiple Targets Multiple Clutters}
In this section we study the scenario with $J$ targets and $K$ clutters. Hence, the incident signal at the surface of the $j$th target is formulated similar to~\eqref{eq:SysModel}. Having multiple targets and multiple clutters motives the study of multi-objective optimization, where the Pareto boundary of the backscattered signal power is investigated. This way, we formulate the weighted sum-backscattered power from the targets under clutters backscattered sum-power constraint as
\begin{subequations}\label{OP:A5}
\begin{align}
\max_{\mathbf{C}}\ & \sum_{j=1}^{J} \gamma_j\text{Tr}\left(\tilde{\mathbf{Z}}_{\text{T}_j}\mathbf{C}_j\right)\tag{\ref{OP:A5}}\\
\quad \text{s.t.}\ & \sum_{k=1}^{K}\text{Tr}\left(\tilde{\mathbf{Z}}_{\text{c}_k}\mathbf{C}\right)\leq \psi,\label{ClutterCons5}\\
&\quad  \eqref{eq:PowerCons1}-\eqref{eq:HermitCons1},
\end{align}
\end{subequations}
where $\gamma_j$ is the weight devoted for the $j$th target. Moreover the backscattered signal sum-power constraint from the clutters is denoted by $\psi$. Solving this problem for a given set of $\gamma_j,\ \forall j\in\mathcal{J}=\{1,\cdots,J\}$ delivers an optimal point on the backscattered power Pareto boundary. Here, the Pareto boundary determines the frontier of the backscattered power region, where an increase in the backscattered power of one target inevitably coincides with a reduction in the backscattered power from at least one other target. For characterizing the Pareto boundary of the backscattered power region, we need to solve problem~\eqref{OP:A4} for multiple combinations of $\boldsymbol{\gamma}=[\gamma_1,\cdots,\gamma_J]$, where $0<\gamma_j<1,\ \forall j\in\mathcal{J}=\{1,\cdots,J\}$, with $|\boldsymbol{\gamma}|_1=1$. 
This problem is a SDP and yields a rank-1 solution~\cite{Palomar2010}. This solution is obtained numerically by interior-point methods. 

\section{Numerical Results}
In this section we discuss the numerical results for the following scenarios,
I) single target at the observation angle $\theta=30$,
II) single target and clutter with position uncertainties, $\theta\in (-50,-20)\cup (20,50)$ and $\phi\in (-20,20)$, respectively,
III) Two targets at the angles $(30,45)$ and three clutters at the angles $(25,36,60)$.
We consider a multilayer structure with three layers. In the simulations, the complex-valued reflection coefficient from deeper boundaries are normalized to the absolute value of reflection coefficient from the surface. Here, we consider that the backscattered signal power from the target(s) is maximized over three and seven time instants, i.e., $N=3, N=7$ jointly. The distance between the antennas are assumed to be half-wavelength, i.e., $D=\frac{\lambda}{2}$.

\begin{figure}
			\centering
				\tikzset{every picture/.style={scale=0.95}, every node/.style={scale=1}}
				\begin{tikzpicture}

\begin{axis}[%
xmin=0,
xmax=1.4,
xlabel={\normalsize $\frac{\text{Backscattered Signal Power of Target 1}}{N}$},
xmajorgrids,
xtick={0,0.2,0.4,0.6,0.8,1,1.2,1.4},
ymin=0,
ymax=1.4,
ylabel={\normalsize $\frac{\text{Backscattered Signal Power of Target 2}}{N}$},
ylabel near ticks,
ymajorgrids,
ytick={0,0.2,0.4,0.6,0.8,1,1.2,1.4},
legend style={at={(axis cs:0.6,0)},anchor= south west, draw=black,fill=white, fill opacity=0.85,legend cell align=left}
]
\addplot [color=black,solid,thick]
  table[row sep=crcr]{0	1.29604784030641\\
  0.986121720783231	1.29604784030641\\
  0.994892023023826	1.29595901077924\\
  1.00360239315819	1.29568938235021\\
  1.012240542314	1.2952345133748\\
  1.02079428805343	1.2945904603989\\
  1.02925163473989	1.29375380464327\\
  1.03760085251164	1.29272167260748\\
  1.04583055351913	1.29149175037567\\
  1.05392976412585	1.29006229135057\\
  1.06188799185676	1.28843211729857\\
  1.0696952860007	1.28660061274909\\
  1.07734229092803	1.28456771295177\\
  1.08482029136112	1.28233388574951\\
  1.09212124902362	1.27990010787186\\
  1.09923783029154	1.27726783628785\\
  1.10616342468568	1.27443897537023\\
  1.11289215428288	1.27141584070206\\
  1.11941887435916	1.26820112039792\\
  1.12573916577767	1.2647978348237\\
  1.13184931978653	1.26120929559143\\
  1.13774631600992	1.25743906467779\\
  1.1434277945089	1.25349091446567\\
  1.14889202285791	1.24936878943706\\
  1.15413785922334	1.24507677015732\\
  1.15916471244154	1.24061904008965\\
  1.16397250007638	1.23599985567011\\
  1.1685616053923	1.23122351996147\\
  1.17293283411565	1.22629436009282\\
  1.17708737177478	1.22121670858414\\
  1.18102674231627	1.21599488855449\\
  1.1847527685927	1.21063320272005\\
  1.18826753521151	1.20513592600669\\
  1.19157335412765	1.19950730153214\\
  1.19467273325903	1.19375153965449\\
  1.19756834830421	1.18787281973868\\
  1.20026301784989	1.18187529425914\\
  1.20275968177188	1.17576309483515\\
  1.20506138285933	1.16954033978461\\
  1.20717125152773	1.16321114278157\\
  1.20909249343318	1.15677962221023\\
  1.21082837975677	1.15024991082578\\
  1.21238223989535	1.14362616535389\\
  1.21375745649093	1.13691257452905\\
  1.21495746151514	1.13011336995738\\
  1.21598573452324	1.12323283405942\\
  1.21684580255573	1.11627530313285\\
  1.21754124002824	1.10924517390715\\
  1.21807566959494	1.1021469069459\\
  1.21845276336546	1.09498502841626\\
  1.21867624423157	1.08776413016493\\
  1.21874988708676	1.08048886973569\\
  1.21874988708676	0\\
  };
\addlegendentry{$\psi=1$}

\addplot [color=red,dash dot,thick]
  table[row sep=crcr]{0 1.14539750160154
  0.831696641200923	1.14539750160154\\
  0.847356836888736	1.14523833966684\\
  0.863252131852327	1.14474578498321\\
  0.879328839070421	1.14389874272056\\
  0.895526644761941	1.14267872080333\\
  0.911778946135365	1.14107060732516\\
  0.928013430459297	1.13906346743667\\
  0.94415294767968	1.13665132106364\\
  0.960116691141696	1.13383385567265\\
  0.975821724014769	1.13061701060305\\
  0.991185011297122	1.12701332725691\\
  1.00612539634569	1.12304210012345\\
  1.02056560380959	1.11872930244775\\
  1.03443611245365	1.11410655526482\\
  1.04767612997698	1.10921055938259\\
  1.06023538164157	1.10408194752378\\
  1.07207621827482	1.09876358303898\\
  1.08317387043283	1.0932991945841\\
  1.09351666164769	1.08773178437889\\
  1.10310557545762	1.08210212385142\\
  1.11194987056328	1.07644964210518\\
  1.12007722240871	1.07080410300318\\
  1.12751330290149	1.06519673268486\\
  1.1342935057609	1.05965162523416\\
  1.14045711152477	1.05418811551572\\
  1.14604481095123	1.04882185098044\\
  1.15109896681998	1.0435636890581\\
  1.15566130460252	1.03842117663448\\
  1.15977232584783	1.03339880506821\\
  1.16347066670953	1.02849851724772\\
  1.16679267199263	1.02372018702562\\
  1.16977214396446	1.01906205604799\\
  1.17244023309606	1.01452111351197\\
  1.17482544268983	1.0100934161976\\
  1.17695368658663	1.00577437520998\\
  1.17884841172656	1.00155897880834\\
  1.18053079223712	0.997441924813917\\
  1.18201982983122	0.993417866557138\\
  1.18333259586319	0.989481417604659\\
  1.18448438983306	0.985627263761347\\
  1.18548891027847	0.981850222405192\\
  1.18635839914657	0.978145385599952\\
  1.18710386065573	0.974507750640489\\
  1.18773509466749	0.970932816057532\\
  1.18826089086384	0.967416213427481\\
  1.18868912377698	0.963953798937407\\
  1.1890268537131	0.96054164997211\\
  1.18928041582966	0.957176057756784\\
  1.18945550034337	0.953853518802843\\
  1.18955722250432	0.950570724120958\\
  1.18959018535975	0.947324595143231\\
  1.18959018535975	0\\
  };
\addlegendentry{$\psi=0.5$}

\addplot [color=green!50!black,dashed,thick]
  table[row sep=crcr]{0	0.252206569296383\\
  0.187189869803791	0.252206569296383\\
  0.19347645646217	0.252142265376405\\
  0.200112879312974	0.251936125844808\\
  0.207120763927234	0.251566304110209\\
  0.21452253819722	0.251008089297844\\
  0.222340495439292	0.250233677773575\\
  0.230595259299266	0.249212098602252\\
  0.239303495637145	0.247909420076458\\
  0.248474755805575	0.246289403442106\\
  0.258107448682137	0.244314794518064\\
  0.268184104883037	0.241949437620928\\
  0.278666404749708	0.239161318205982\\
  0.289490779155907	0.235926472521293\\
  0.300565675730008	0.232233446225755\\
  0.311771748984402	0.228087652169577\\
  0.322965985626777	0.223514703459304\\
  0.333989993103828	0.218561759790161\\
  0.344681746443609	0.21329613718032\\
  0.354888973176052	0.207801021365195\\
  0.3644816374811	0.202168913694057\\
  0.373361286063902	0.196493970047969\\
  0.381465578312716	0.190864810169397\\
  0.388768078937936	0.185358850141704\\
  0.395273903835223	0.180039040715092\\
  0.401013053974183	0.174952872750845\\
  0.406035153558683	0.170131076574668\\
  0.410403808206792	0.165587407914103\\
  0.414178036495263	0.161334454091642\\
  0.41742651198132	0.157367054440068\\
  0.420212115626506	0.153677295442152\\
  0.422592948601559	0.150253840778344\\
  0.424623973197375	0.147079674565788\\
  0.426353143863486	0.144137790167319\\
  0.427821745916558	0.141412551174097\\
  0.429066430841093	0.138887445128909\\
  0.430119133161749	0.136546167755712\\
  0.431007294782589	0.134373413315709\\
  0.431754451817131	0.132354927936903\\
  0.432380753820271	0.130477539975948\\
  0.432903428389867	0.128729162035294\\
  0.43333720021991	0.127098742635936\\
  0.433694649488314	0.125576227520511\\
  0.433986529813802	0.124152487896979\\
  0.43422204083226	0.122819244111648\\
  0.434409059697941	0.121568995665782\\
  0.43455433876028	0.120394951095612\\
  0.434663673471263	0.119290961104942\\
  0.434742044750075	0.118251455461673\\
  0.434793739448282	0.117271384465344\\
  0.434822448020176	0.116347171346366\\
  0.434831388634239	0.115472598671609\\
  0.434831388634239	0\\
  };
\addlegendentry{$\psi=0.1$}

\end{axis}
\end{tikzpicture}%
\caption{Backscattered signal power region of two targets at $(30,45)$ in the presence of three clutters at $(25,36,60)$. $N=3$ and $M=4$.}
\label{fig:Region}
\end{figure}
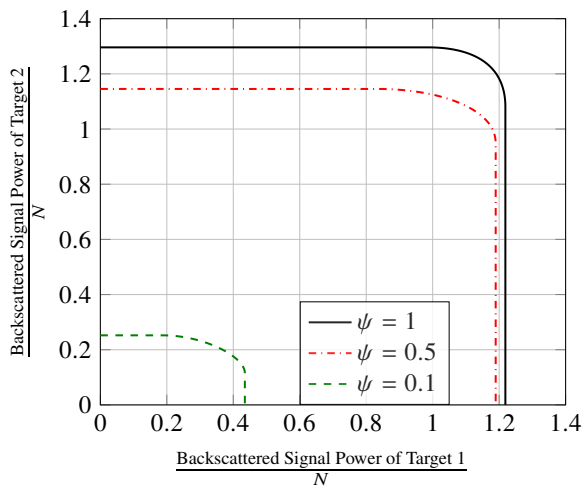

Provided the targets observation angle $\theta=30$ and the target's multilayer material (scenario I), the backscattered signal power from the target can almost be doubled compared to the case with unknown material. This can be seen from Fig.~\ref{fig:SingleTarget}. Moreover, by increasing the temporal dimensions in the waveform design process (from $N=3$ to $N=7$), we can save in the number of spatial dimensions (antennas) significantly. As can be seen from Fig.~\ref{fig:SpatialTemporalTradeoff} solid curve, we obtain equal backscattered signal power with $(M,N)=(3,7)$ and $(M,N)=(10,3)$. Inaccuracy in target positioning yields in an extreme backscattered signal power reduction from the angle of interest. As can be seen from Fig.~\ref{fig:SingleTarget}, $20$ degrees inaccuracy in positioning results in almost $80\%$ reduction in backscattered signal power. Hence, we studied a robust design problem including a target in an inaccuracy region of $\theta\in (-50,-20)\cup (20,50)$. Moreover, the backscattered signal power in $\phi\in (-20,20)$ is kept low, i.e., $\xi\rightarrow 0$, due to the presence of a clutter. In Fig.~\ref{fig:SingleTargetCoverRegionWithClutterCoverRegion}, we observe the superiority of spatio-temporal waveform with known target's material compared with the unknown case from the backscattered signal power perspective. Considering two objects and three clutters, the Pareto boundary of the backscattered signal power is delivered through weighted sum maximization. By inducing the sum backscattered signal power constraint from the clutters, i.e., $\psi$, the optimal solutions stay on the Pareto boundary. This boundary shrinks by decreasing $\psi$, which is depicted in~Fig.~\ref{fig:Region}. Intuitively, for sufficiently low $\psi$, optimal transmit waveform becomes spatio-temporal null-steering in the direction of the clutters.
\section{Conclusion}
In this paper we exploited the targets response to the incident signal for backscattered signal power maximization. The spatio-temporal waveform design for the application of single target and multiple targets with multiple clutters falls into the category of polynomial-time solvable problems. Hence, the backscattered signal power maximization problems for both cases are formulated as convex programs. These problems are essentially semidefinite programs (SDP). The first problem is analytically solved, however the second and third problems are solved numerically. Finally, we observe that, leveraging the targets response in spatio-temporal waveform design yields in almost doubling the backscattered signal power from the targets surface. The backscattered signal power can be even further enhanced by increasing $N$ in the precoding phase. Moreover, we observed that by increasing the number of temporal dimensions in the waveform design phase, we can save in the number of antennas in the transmission phase.
\section{Acknowledgment}
This work was supported by the German Research
Foundation (DFG) for the CRC/TRR 196 MARIE.

\bibliographystyle{IEEEtran}
\bibliography{reference}
\end{document}